\documentstyle[epsfig,prb,aps]{revtex}
\begin{document}

\title{Theory for the electromigration wind force in dilute alloys}

\author{J.P. Dekker and A. Lodder}

\address{Faculteit Natuurkunde en Sterrenkunde, Vrije Universiteit,
         De Boelelaan 1081, 1081 HV Amsterdam, The Netherlands}

\author{J. van Ek}

\address{Department of Physics, Tulane University,
         New Orleans, LA 70118}

%

\date{\today}
\maketitle

\begin{abstract}
\normalsize{
A multiple scattering formulation for the electromigration wind force on
atoms in dilute alloys is
developed.
The theory describes electromigration via a
vacancy mechanism.
The method is used to calculate the wind valence for electromigration
in various host metals having a close-packed lattice structure,
namely aluminum, the noble metals copper, silver and gold and
the $4d$ transition metals.
The self-electromigration results for aluminum and the noble metals
compare well with experimental data.
For the $4d$ metals small wind valences are found, which make these
metals attractive candidates for the
experimental study of the direct valence.
}
\end{abstract}

\section{Introduction}
\label{sec:introduction}

An electric field, applied to a metal sample, causes an atomic current,
besides the common electric current.
This process is called electromigration and has been studied for many
years.
Already in 1930 Coehn and Specht \cite{CoeSpe30} reported a flow of
hydrogen atoms towards the cathode in palladium.
This migration of interstitial atoms is a rather harmless manifestation
of the effect.
It offers the possibility to influence the concentration profile of
impurity atoms along a sample.
However, a net flow of host atoms can also be the consequence of
applying an electric field.
Clearly, this reduces the conductivity of a metal wire
because of the formation of voids on one side and hillocks on the other 
side.
In wires of normal size this turns out not to be of great importance,
because the atoms do not move very fast.
In thin films, however, the situation is different.
Thin films can carry higher current densities, because the heat, produced
by the current, is transfered more easily to the environment.
Therefore they can endure a current density greater than
$~10^{7}{\rm A}/{\rm cm}^{2}$, while a bulk sample would melt when the current
density is $~10^{4}{\rm A}/{\rm cm}^{2}$.
Electromigration damage in aluminum films is a well-known example.

The migration of atoms turns out to be influenced by the presence of
impurities.
Most important is the effect of copper doping of Al films,
which greatly reduces the electromigration damage.
This effect is known for a long time without understanding it, so
impurities have been added using the method of trial and error.
Furthermore, although the damage appears too fast for a device user,
it is still not fast enough for an experimental physicist.
So, the effect is studied under so-called accelerated conditions, high
temperatures and high current densities.
Unfortunately, the extrapolation to user conditions is not trivial,
because the total atomic flow is a result of a number of
competing contributions with different activation energies.
All of this complicates the picture and cries for a theoretical framework,
to which this article is aimed to contribute.

Two ingredients are essential in the process of electromigration,
diffusion and a driving force.
The motion of atoms due to diffusion is random.
At lower temperatures diffusion along grain boundaries is dominant,
but at high temperatures bulk diffusion becomes the most important
contribution.
For accelerated conditions it is the leading mechanism
and the atoms move by exchanging positions with
vacancies.
The second ingredient, the driving force, introduces a bias
into the random motion of atoms.
One contribution to that force is calculated in the present work,
namely the wind
force due to scattering of electrons by the moving atom, which can be
either a host atom or an impurity atom.
A well-established quantum-mechanical expression for the wind force is
available \cite{SorLodHov82}.
Besides the wind force, the driving force
has a direct part.
If the nuclear charge is not completely screened, the electric field 
directly pushes the atom towards the cathode.
This force has been the subject of a controversy for many years
\cite{HoKwo89,Lod91clar}.
Both contributions are proportional to the electric field and the total
driving force can therefore be written as
\begin{equation}
\label{eq:Ftot}
{\bf F}={\bf F}_{\rm wind}+{\bf F}_{\rm direct}=(Z_{\rm wind}+Z_{\rm direct})
e{\bf E}=Z^{*}e{\bf E}.
\end{equation}
The effective valence $Z^{*}$, which is the sum of the wind and direct
valence, has been measured for a lot of systems \cite{HoKwo89}.
Often the wind force dominates the direct force and, depending on the
system, the resulting total force can push the atom either to the anode or
to the cathode.

The calculation of the wind force requires knowledge of the
electronic structure of the alloy.
An impurity in a dilute alloy, in which the concentration of impurities
is low, only interacts with its direct environment of host atoms.
The electrons feel a potential, which only differs from the host
potential in a small cluster of atoms containing the migrating atom, the
vacancy and the surrounding host atoms affected by charge transfer
and lattice deformation.
A Green's function formulation of multiple scattering theory
is used to calculate the electron wave function in the dilute alloy.
This formalism is an extension of the
Korringa-Kohn-Rostoker (KKR) method for the calculation of the
band structure of a metal \cite{k,kr}.
The alloy is described with respect to a reference system.
The conventional choice of this reference system is
the host metal, but then the electronic structure of the
important saddle point configuration in electromigration
cannot be described.
This is due to the fact, that
the atoms in the alloy system cannot be mapped onto the ones in
the host system one-to-one.
As proposed by Lodder \cite{Lod76} an intermediate void system is used,
which does not contain any scatterers in the perturbed region.
The original formulation \cite{1984}, in which the cluster
of perturbed atoms
is considered as a whole, was recently applied in a preliminary
study of electromigration in copper and aluminum \cite{VEkDekLod95}.
This formulation is rather cumbersome and in the mean time we found
a simpler single site formulation,
in which each atom in the cluster is considered separately,
analogous to the formalism used for the description of interstitial
impurities by Oppeneer \cite{OppLod87I} and van Ek \cite{EkLod91I}.
The validity of this method is not limited to the
particular problem of substitutional electromigration.
In all problems concerning dilute alloys
the wave function can be calculated this way.

The multiple scattering expression for the wind force is elaborated in
section \ref{sec:expression}.
This section contains two subsections, devoted to
the wave function and the Green's function matrix respectively.
In section \ref{sec:calc} some details of the calculation are given.
The formalism is applied to metals with a close-packed lattice structure
in section \ref{sec:results}.
Self-electromigration and impurity electromigration in Al are
treated in subsection \ref{subsec:Al}, self-electromigration in
the noble metals as well as impurity migration in Ag
in subsection \ref{subsec:noble} and self-electromigration in
$4d$ transition metals in subsection \ref{subsec:4dtrans}.
In section \ref{sec:conclusions} the main results of this paper are
summarized.
 
Throughout the article atomic units are used, such that
$\hbar = 2m =1$.
Exceptions are stated explicitly.

\section{Multiple-scattering expression for the wind force}
\label{sec:expression}

In this section the quantum-mechanical equation for the
wind force of Sorbello et al.
\cite{SorLodHov82}
\begin{equation}
\label{eq:WFexpr}
{\bf F}_{{\rm wind}}=\sum_{k} \delta f(k)
<\Psi_{k}|-\nabla_{{\bf R}_{p}}v_{p}|\Psi_{k}>,
\end{equation}
which traces back to the pioneering work of Bosvieux and Friedel
\cite{BosFri62},
and in which
\begin{equation}
\label{eq:deltaf}
\delta f(k)=e \tau_{k}{\bf E}\cdot {\bf v}_{k}
df_{0}(\epsilon_{k})/d\epsilon_{k},
\end{equation}
is expressed in computable quantities.
In Eq. (\ref{eq:WFexpr}) the alloy electron wave function, $\Psi_{k}$,
is supposed to be
constructed from the corresponding Bloch function, labeled by
crystal momentum ${\bf k}$ and band index $n$,
combined in $k=({\bf k},n)$, and having an energy eigenvalue
$\epsilon_{k}$.
In a simple ballistic picture momentum is transfered from electrons
to the atom
because of scattering, resulting in the wind force.
The momentum transfered to this atom, with a potential
$v_{p}$ centered at position ${\bf R}_{p}$ in the alloy,
is proportional to the expectation value of the force operator
$-\nabla_{{\bf R}_{p}}v_{p}$.
When no electric field is present, the electrons
are distributed according to the Fermi-Dirac function
$f_{0}(\epsilon_{k})$, leading to an exact cancellation of forces.
In other words, each push from the electrons in one direction is
compensated, on the average, by one in the opposite direction.
This symmetry is broken by the presence of the electric field, which
alters the distribution function by $\delta f(k)$.
The first and leading term of this deviation, Eq. (\ref{eq:deltaf}), is
linear in the applied electric field ${\bf E}$ and proportional to
the electron transport relaxation time $\tau_{k}$.
In Eq. (\ref{eq:deltaf})
$e$ is the elementary charge and ${\bf v}_{k}$ is the velocity of the
electron.
As usual, in all applications, the transport relaxation time is taken
constant throughout the Brillouin zone.
All other quantities can be extracted from the electronic structure.
The method for the calculation of the electron wave function
originally provided
by Lodder \cite{1984} will be used in a modified way.

In the present formulation the crystal is divided into cells.
For the host crystal these cells
are the well-known Wigner-Seitz cells containing
local potentials $v^{{\rm h}}({\bf x})$.
For the alloy these cells may have a different shape in the region of
the impurity cluster and they contain local potentials $v^{p}({\bf x})$.
The local potentials $v^{{\rm h}}$ and $v^{p}$ consist of a spherical
atomic potential surrounded by an interstitial
region, where the potential is constant.
In order to evaluate the matrix element appearing in Eq. (\ref{eq:WFexpr}),
the wave function is expanded in the $p$-th cell, centered at ${\bf R}_{p}$.
In such a cell the wave function
$\Psi_{k}({\bf r})=\Psi_{k}({\bf x}+{\bf R}_{p})$
can be written as a linear combination of regular solutions
$R_{L}^{p}$ of the Schr\"odinger equation with local potential $v^{p}$, so
\begin{equation}
\label{eq:localPsi}
\Psi_{k}({\bf x}+{\bf R}_{p})=\sum_{L}c_{kpL}R_{L}^{p}({\bf x}),
\end{equation}
where $L=(l,m)$ combines the angular momentum and magnetic quantum numbers.
The expansion coefficients $c_{kpL}$
will be derived in Subsection \ref{subsec:wave}.
The basis functions $R_{L}^{p}$ are constructed such that outside the
atomic sphere they have the free space form
\begin{equation}
\label{eq:RLp}
R_{L}^{p}({\bf x})=j_{L}({\bf x})-i \sum_{L'}
t^{p}_{LL'}h_{L'}^{+}({\bf x}),
\end{equation}
where $j_{L}({\bf x})$ is the product of a spherical Bessel function
$j_{l}(\kappa x)$ and a spherical harmonic $Y_{L}({\hat x})$,
$h_{L}^{+}({\bf x})$ is a similar product for the spherical Hankel function
$h_{l}^{+}(\kappa x)$, $\kappa={\sqrt E}$ and
$t^{p}$ is the scattering matrix for the potential $v^{p}$
embedded in free space.
For a spherically symmetrical scatterer this matrix is diagonal and equals
$t^{p}_{LL'}=-\sin \eta_{l}^{p} e^{i \eta_{l}^{p}}\delta_{LL'}$,
the $\eta$ being the phase-shifts.

In that case the wind force expression (\ref{eq:WFexpr}) can be elaborated
to the form
\begin{equation}
\label{eq:expval4}
{\bf F}_{{\rm wind}}=\sum_{k} \delta f(k)
2 {\rm Re} \sum_{L}\sum_{m_{1}=-(l+1)}^{l+1}
c^{*}_{kpL}{\bf D}_{L;l+1,m_{1}}\sin(\eta_{l+1}-\eta_{l})
e^{i(\eta_{l+1}-\eta_{l})}c_{kpl+1,m_{1}},
\end{equation}
according to Eqs. (26) and (31) in Ref. \cite{1984},
in which the vectorial matrix ${\bf D}$ is defined as
\begin{equation}
\label{eq:vectGaunt}
{\bf D}_{LL'}=\int d{\hat x} Y_{L}({\hat x}){\hat x}Y_{L'}({\hat x}).
\end{equation}

\subsection{The wave function coefficients}
\label{subsec:wave}

\begin{figure}[htb]
\epsfig{figure=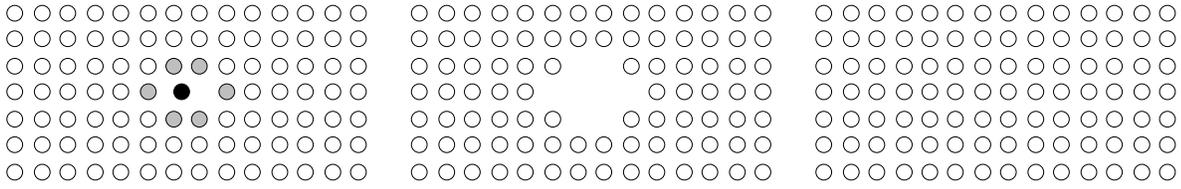,height=2.35cm}
\caption{The systems used in the description of the dilute alloy (left): the
void system (center) and the host system (right).}
\label{fig:void}
\end{figure}

The alloy wave function coefficients that appear in
Eq. (\ref{eq:localPsi}) will be expressed in terms of wave function
coefficients $c^{\rm host}_{kjL}$,
which appear in the local expansion of Bloch function
$\Psi_{k}^{{\rm host}}({\bf r})=\Psi_{k}^{{\rm host}}({\bf x}+{\bf R}_{j})$,
\begin{equation}
\label{eq:Psih}
\Psi^{\rm host}_{k}({\bf x}+{\bf R}_{j})=
\sum_{L}c^{\rm host}_{kjL}R_{L}^{j}({\bf x}),
\end{equation}
where $R_{L}^{j}$ are the regular solutions of the Schr\"odinger
equation within the Wigner-Seitz cell labeled by $j$
and contain $t^{{\rm h}}$ instead of $t^{p}$ in their asymptotic form
(see Eq. (\ref{eq:RLp})).
Unperturbed atomic positions ${\bf R}_{j}$ in the host do not necessarily
coincide with one of the atomic positions ${\bf R}_{p}$ in the alloy.
The number of alloy sites may even differ from the number of host
lattice sites.
Therefore, although the alloy is described with respect to the host, the host
system cannot be used as a reference system straightforwardly.
An intermediate system is required, which will be referred to as
the void system.
This system consists of a void region, where the potential is constant,
surrounded by host atom potentials.
The void region is chosen to be as extended as the perturbed region
in the dilute alloy.
This system can serve as a
reference system for both the host and the alloy, as it
is represented symbolically in Fig. \ref{fig:void}.
The relation between the host and alloy wave function is given via
the void wave function.
The alloy wave function is expressed in terms of
the void wave function by the
Green's function expression \cite{1984},
\begin{equation}
\label{eq:GvE}
\Psi_{k}=\Psi_{k}^{{\rm void}}+G^{{\rm void}} \sum_{p}v^{p}\Psi_{k}.
\end{equation}
Analogously to Eqs. (\ref{eq:localPsi}) and (\ref{eq:Psih}) the
void wave function can be written in terms of local basis
functions, being spherical Bessel functions because of the
constant potential in the void region
\begin{equation}
\label{eq:Psivoid}
\Psi^{\rm void}_{k}({\bf x}+{\bf R}_{p})=
\sum_{L}c^{\rm void}_{kpL}j_{L}({\bf x}).
\end{equation}
Inside the void, the void Green's function can be written as
\begin{eqnarray}
\label{eq:Gvoid}
&&G^{{\rm void}}({\bf x}+{\bf R}_{p},{\bf x}'+{\bf R}_{p'})=
-i\kappa \sum_{L}j_{L}({\bf x}_{<})h^{+}_{L}({\bf x}_{>})
\delta_{pp'}\nonumber \\
&&+\sum_{LL'}j_{L}({\bf x})
{\cal G}^{{\rm void},pp'}_{LL'}j_{L}({\bf x}').
\end{eqnarray}
This expression with the void Green's function matrix
${\cal G}^{{\rm void},pp'}$ is a straightforward generalization
of the expression for the Green's function of free space
\begin{eqnarray}
\label{eq:G0}
&&G^{0}({\bf x}+{\bf R}_{p},{\bf x}'+{\bf R}_{p'})=
-i\kappa \sum_{L} j_{L}({\bf x}_{<})h_{L}^{+}({\bf x}_{>})\delta_{pp'}+
\nonumber \\
&&\sum_{LL'}j_{L}({\bf x})B^{pp'}_{LL'}j_{L'}({\bf x}'),
\end{eqnarray}
in which the free space propagation matrix $B$ appears, defined as
\begin{equation}
\label{eq:B}
B^{pp'}_{LL'}\equiv B_{LL'}({\bf R}_{pp'})=
4 \pi i^{l-l'-1}\kappa (1-\delta_{pp'})
\sum_{L''}i^{l''}C_{LL'L''}h^{+}_{L''}({\bf R}_{pp'}).
\end{equation}
The void Green's function matrix will be given in Subsection \ref{subsec:Gvoid}.

Combining Eqs. (\ref{eq:GvE}), (\ref{eq:Psivoid}) and (\ref{eq:Gvoid})
the following matrix equation can be found
\begin{equation}
\label{eq:coefp}
c^{{\rm void}}_{kpL}=
\sum_{p'L'}(1-{\cal G}^{{\rm void}}t)^{pp'}_{LL'}c_{kp'L'}.
\end{equation}
Details of the derivation will be given in Appendix \ref{app:B}.
Following the same procedure, void wave function coefficients carrying
the host position label $j$
can be expressed in terms of host coefficients $c^{{\rm host}}_{kjL}$
\begin{equation}
\label{eq:coefj}
c^{{\rm void}}_{kjL}=\sum_{j'L'}(1-{\cal G}^{{\rm void}}t^{h})^{jj'}_{LL'}
c^{{\rm host}}_{kj'L'}.
\end{equation}
Since $c^{{\rm void}}_{kpL}$ and $c^{{\rm void}}_{kjL}$ are related by
\begin{equation}
\label{eq:cvjcvp}
c^{{\rm void}}_{kpL}=
\sum_{L_{1}}J^{pj}_{LL_{1}}c^{{\rm void}}_{kjL_{1}},
\end{equation}
Eqs. (\ref{eq:coefp}) and (\ref{eq:coefj}) can be combined to
\begin{eqnarray}
\label{eq:coef}
&&\sum_{p'L'}
(1-{\cal G}^{{\rm void}}t)^{pp'}_{LL'}c_{kp'L'}=\nonumber \\
&&\sum_{j'L'L_{1}}J^{pj}_{LL_{1}}
(1-{\cal G}^{{\rm void}}t^{{\rm h}})^{jj'}_{L_{1}L'}
c^{{\rm host}}_{kj'L'}.
\end{eqnarray}
It is noteworthy that in this equation two types of summations
over the angular momentum occur.
The first type has a natural cut-off because of the multiplication by
$t$-matrices, and the second type, such as the summation over $L_{1}$,
in principle runs to infinity.
The latter summation
can be treated analytically, to be shown in Appendix \ref{app:C},
and the final equation used
in the actual calculations,
\begin{eqnarray}
\label{eq:coef1}
&&\sum_{p'}(1-{\cal G}^{{\rm void}}t)^{pp'}c_{kp'}=\nonumber \\
&&b({\bf k},{\bf R}_{pj})e^{i{\bf k}\cdot{\bf R}_{pj}}
t^{{\rm h}}c^{{\rm host}}_{kj}
-\sum_{j'}{\cal G}^{{\rm void},pj'}t^{{\rm h}}c^{{\rm host}}_{kj'},
\end{eqnarray}
only contains summations with a natural cut-off.
Angular momentum labels have been dropped in order to simplify the
notation.
The matrix $b({\bf k},{\bf R})$ is defined by
\begin{equation}
\label{eq:bkp}
b({\bf k},{\bf R})=
\sum_{j'}B({\bf R}-{\bf R}_{j'})e^{-i{\bf k}\cdot({\bf R}-{\bf R}_{j'})},
\end{equation}
in which the summation runs over {\it all} lattice sites.

The host wave function coefficients occurring in Eq. (\ref{eq:coef1})
are calculated by the KKR method according to
\begin{equation}
\label{eq:KKRc}
\sum_{L'}M_{LL'}({\bf k})t^{h}_{l'}c^{\rm host}_{kjL}=0.
\end{equation}
This equation contains the KKR matrix
\begin{equation}
\label{eq:KKRmat}
M({\bf k})={t^{h}}^{-1}-b({\bf k}),
\end{equation}
in which $b({\bf k})=b({\bf k},0)$, defined by Eq. (\ref{eq:bkp}).

In its original form the KKR method applies to muffin-tin
potentials.
However, as shown by Nesbet \cite{Nes90}, it is possible to circumvent
this restriction on the crystal potential in the same way
as has been done in the derivation of Eq. (\ref{eq:coef}).
Now we will use this procedure in the calculation of
the void Green's function.

\subsection{The void Green's function}
\label{subsec:Gvoid}

The void Green's function and the host Green's function $G^{{\rm h}}$ are
related by the Lippmann-Schwinger equation
\begin{equation}
\label{eq:GvisGh}
G^{{\rm void}}=G^{{\rm h}}-G^{{\rm h}}\sum_{j}v_{j}^{{\rm h}}G^{{\rm void}}.
\end{equation}
The summation runs over the positions in the void.
This equation is again evaluated in the position representation.
The host Green's function is given by
\begin{eqnarray}
\label{eq:Gh}
&&G^{{\rm h}}({\bf x}+{\bf R}_{j},{\bf x}'+{\bf R}_{j'})=
\sum_{L}R_{L}({\bf x}_{<})S_{L}({\bf x}_{>})
\delta_{jj'}\nonumber \\
&&+\sum_{LL'}R_{L}({\bf x}){\cal G}^{jj'}_{LL'}R_{L}({\bf x}').
\end{eqnarray}
Outside the atomic sphere the regular solutions $R_{L}$ of the local
Schr\"odinger equation have the asymptotic form of Eq. (\ref{eq:RLp})
with $t^{p}$ replaced by $t^{{\rm h}}$
and the singular solutions $S_{L}$ equal $-i \kappa h^{+}_{L}$.
First the void Green's function matrix ${\cal G}^{{\rm void},jj'}$,
carrying host position labels $j$ and $j'$, will be
expressed in terms of the
host Green's function matrix ${\cal G}^{jj'}$, to be determined later.
To that end Eq. (\ref{eq:GvisGh}) is elaborated at the positions
${\bf x}+{\bf R}_{j}$ and ${\bf x}'+{\bf R}_{j'}$,
where both ${\bf x}$ and ${\bf x}'$ are located close to their cell centers
${\bf R}_{j}$ and ${\bf R}_{j'}$ respectively.
Following similar steps as in Appendix \ref{app:B} one finds the
relation
\begin{equation}
{\cal G}^{{\rm void},jj'}={\cal G}^{jj'}
-\sum_{j_{1}}{\cal G}^{jj_{1}}t^{{\rm h}}
{\cal G}^{{\rm void},j_{1}j'}
\label{eq:GVisGminGAG}
\end{equation}
where the angular momentum labels have been dropped.
The matrix $t^{{\rm h}}$ is the same at all lattice
sites, so the site label has been dropped.
Iteration of this equation leads to
\begin{equation}
{\cal G}^{{\rm void},jj'}={\cal G}^{jj'}
-\sum_{j_{1}j_{2}}{\cal G}^{jj_{1}}
{({t^{{\rm h}}}^{-1}+{\cal G})}^{-1}_{j_{1}j_{2}}
{\cal G}^{j_{2}j'}.
\label{eq:GVisGminGAG2}
\end{equation}

The host Green's function matrix ${\cal G}$
is calculated using the Lippmann-Schwinger equation
\begin{equation}
\label{eq:GhinG0}
G^{{\rm h}}=G^{0}+G^{0}\sum_{j}v_{j}^{{\rm h}}G^{{\rm h}},
\end{equation}
$G^{0}$ being the free-space Green's function.
The summation runs over all host lattice sites.
Working in the position representation and using Eqs. (\ref{eq:G0}) and
(\ref{eq:Gh}) one finds the following matrix equation
\begin{equation}
\label{eq:calG0}
{\cal G}^{jj'}=B^{jj'}+\sum_{j_{1}} B^{jj_{1}}t^{{\rm h}}{\cal G}^{j_{1}j'}.
\end{equation}
Taking the Fourier transform of this equation one finds
\begin{equation}
\label{eq:calGk}
{\cal G}({\bf k})\equiv
\sum_{j'}{\cal G}^{jj'}e^{i{\bf k}\cdot {\bf R}_{jj'}}=
b({\bf k})+b({\bf k})t^{{\rm h}}{\cal G}({\bf k}).
\end{equation}
This equation can be solved for ${\cal G}({\bf k})$
\begin{equation}
\label{eq:calGk1}
{\cal G}({\bf k})={(1-b({\bf k})t^{{\rm h}})}^{-1}b({\bf k})=
b({\bf k})+b({\bf k})M^{-1}({\bf k})b({\bf k}),
\end{equation}
containing the KKR matrix, Eq. (\ref{eq:KKRmat}).
For the reverse Fourier transform,
using the right hand side of this equation,
one obtains
\begin{equation}
\label{eq:calG}
{\cal G}^{jj'}=\frac{1}{\Omega_{\rm BZ}}\int_{\rm BZ}
d^{3}k \left[b({\bf k})+b({\bf k})M^{-1}({\bf k})b({\bf k})\right]
e^{i{\bf k}\cdot{\bf R}_{jj'}},
\end{equation}
in which the ${\bf k}$ summation has been replaced by an integral over
the Brillouin zone (BZ) with volume $\Omega_{\rm BZ}$.

By now, expressions for ${\cal G}^{{\rm void}}$ and ${\cal G}$ have
been derived for centers $j$ of host cells only.
Interestingly, the generalized relation
\begin{equation}
\label{eq:GV}
{\cal G}^{{\rm void},nn'}={\cal G}^{nn'}-
\sum_{j_{1}j_{2}}{\cal G}^{nj_{1}}
{({t^{{\rm h}}}^{-1}+{\cal G})}^{-1}_{j_{1}j_{2}}
{\cal G}^{j_{2}n'}
\end{equation}
turns out to be valid also,
carrying arbitrary position labels $n$ and $n'$,
which can refer to host as well as alloy positions.
The host Green's function matrix ${\cal G}^{nn'}$ is calculated from
\begin{equation}
\label{eq:calGnn}
{\cal G}^{nn'}\equiv \frac{1}{\Omega_{BZ}}\int_{BZ}d^{3}k
[b({\bf k},{\bf R}_{nn'})+b({\bf k},{\bf R}_{n})M^{-1}({\bf k})
b^{{\rm T}}(-{\bf k},{\bf R}_{n'})]e^{i{\bf k}\cdot{\bf R}_{nn'}},
\end{equation}
in which the matrix $b({\bf k},{\bf R})$ is defined in Eq. (\ref{eq:bkp}).
The derivation is given in Appendix \ref{app:C}.

Eq. (\ref{eq:coef1}), together with Eqs. (\ref{eq:GV}) and (\ref{eq:calGnn}),
are the key equations of the theory presented.
The alloy wave function coefficients $c_{kpL}$ can be calculated, if the
host wave function coefficients
$c^{{\rm host}}_{kjL}$ and the matrix ${\cal G}^{{\rm void}}$ are known.
The host wave function coefficients can be calculated using the
KKR equation (\ref{eq:KKRc}).
The matrix ${\cal G}^{{\rm void}}$ can be calculated after evaluation of the
Brillouin zone integral in Eq. (\ref{eq:calGnn}).

\section{Computational details}
\label{sec:calc}

In this section a few quantities which are necessary in the calculation
are discussed.
As mentioned in Sec. \ref{sec:expression} the transport relaxation time
$\tau$ is estimated from the measured resistivity.
The procedure is given in Subsection \ref{subsec:tau}.
The construction of the potentials is discussed in Subsection
\ref{subsec:v}.
Finally, the procedure for the extraction of
measurable quantities from the calculations
is explained in Subsection \ref{subsec:comparison}.

\subsection{The transport relaxation time $\tau$}
\label{subsec:tau}

Upon accepting the widely used isotropic transport
relaxation time approximation for $\tau$,
the bulk conductivity (or the inverse bulk
resistivity) for cubic crystals is given by \cite{Zim72}
\begin{equation}
\sigma=\frac{1}{\rho}=\frac{e^{2}\tau}{4{\pi}^{3}\hbar}\frac{1}{3}
\int_{FS}dS_{k}v_{k}.
\label{eq:tau}
\end{equation}
The integral of the electron velocity $v_{k}$ over the Fermi surface is
directly computable from the electronic structure of the host metal.
The only quantity left is the transport relaxation time $\tau$, to which
the wind force is directly
proportional as can be seen from Eqs.
(\ref{eq:WFexpr}) and (\ref{eq:deltaf}).
Combining Eqs. (\ref{eq:Ftot}), (\ref{eq:WFexpr}),
(\ref{eq:deltaf}) and (\ref{eq:tau}),
the temperature dependent wind valence can be written as
\begin{equation}
Z_{{\rm wind}}(T)=\frac{K(T)}{\rho(T)}.
\label{eq:K}
\end{equation}
The temperature dependence of $K$ comes from the temperature dependence of
the Fermi-Dirac function appearing in Eq. (\ref{eq:deltaf}) and
can be neglected up to the melting temperature of most metals.
The remaining temperature dependence of the wind valence through the 
resistivity is well-known
from the ballistic model by Fiks \cite{Fik59} and Huntington and Grone
\cite{HunGro61} and from later more sophisticated models
\cite{KumSor75}.

It should be noted that the calculated transport relaxation time can
differ considerably from the free electron
transport Drude relaxation time, ${\tau}^{\rm FE}$, arising from the equation
\begin{equation}
\label{eq:taufe}
\sigma=\frac{1}{\rho}=\frac{ne^{2}{\tau}^{\rm FE}}{m}.
\end{equation}
$\tau$ can be as much as a factor of 17 larger than ${\tau}^{\rm FE}$,
as was found for palladium \cite{EkLod91II}.

\subsection{Potentials}
\label{subsec:v}

Both the host and alloy potentials are modelled by muffin-tin
potentials \cite{MolLodCol83}.
The corresponding phase shifts, {\it c.q.} the $t$ matrices $t^{\rm h}$
and $t^{p}$, serve as input for the computational
procedure.
The alloy potentials are constructed
only for two positions
of the migrating atom: the initial position and the position half-way,
the saddle point.
When the atom is somewhere else along the path,
which runs along a straight line in the $<110>$ direction,
the phase shifts are calculated using the interpolation formula
\begin{equation}
\label{eq:etainterp}
\eta_{l}^{p}(s)=\cos^{2}({\pi s})\eta_{l}^{p}(0)+
\sin^{2}({\pi s})\eta_{l}^{p}(\frac{1}{2}).
\end{equation}
The variable $s$ runs from 0 to 1 along the path, so
$\eta_{l}^{p}(0)=\eta_{l}^{p,{\rm initial}}$ and halfway
$\eta_{l}^{p}(\frac{1}{2})=\eta_{l}^{p,{\rm saddle point}}$,
where $p$ refers to one of the perturbed atoms in the cluster or
to the migrating atom.
The $s$ dependence of the interpolation formula guarantees a smooth
behaviour of the phase shifts at the saddle point.

A parameter, which influences the results of the calculation, is the
muffin-tin radius.
The choice of this parameter is bound by two conflicting conditions.
On the one hand it must be small enough to leave some room for the migrating
atom, while on the other hand it must be large enough such that the
potential still reproduces the electronic properties of the metal
as much as possible.
For FCC metals a muffin-tin radius of $0.325 a$ is an optimal
value \cite{OppLod87II}.

A real physical system is charge neutral.
This can be established using a generalized Friedel sum rule \cite{LodBra94}
and a shifting procedure proposed by Lasseter and Soven \cite{LasSov73}
and successfully applied in numerous systems.
The potential of the atom is shifted by a constant value,
which can be interpreted as the addition of charge on the muffin tin sphere.
For the different configurations during the jump various choices are
possible for the shift procedure.
If the migrating atom is in its initial position, the simplest choice is a
shift of the potentials
of all nearest neighbours of the vacancy by the same value.
In the saddle point configuration all nearest neighbours of the two small
moon-shaped vacancies surrounding the migrating atom can be shifted.
We want to point out that in the latter configuration
the small vacancies are not accounted for, which
means that the shifting procedure also must correct for the
corresponding loss of charge in that region of space.
This makes it unclear, what the shifting procedure means for the 
electronic properties.
Due to these uncertainties and
the lack of self-consistent potentials we therefore have decided
to use the phase shifts corresponding to the unshifted potentials.
Test calculations show, that the potential shift leads to a maximum
change in the wind valence of about
10 \% and in most cases of only a few per cent.
In view of the limited accuracy of most measurements this is acceptable
for the time being.
In principle full-potential calculations can reveal detailed information
of the charge state of the impurity cluster.

\subsection{Comparison with experiment}
\label{subsec:comparison}

The calculated wind valence is a position dependent second-rank
tensor.
In order to compare with experiment the tensor has to be reduced to a scalar.
The relevant
component of the force is the one in the direction of the migration
path, which we indicate by $\hat{s}$.
Averaging over all orientations of the lattice, simulating the
polycrystalline samples used in practice,
we yield for the scalar wind valence
\begin{equation}
\label{eq:dirav}
Z_{\rm wind}=\hat{s}^{T}\cdot \underline{Z_{\rm wind}}\cdot \hat{s}.
\end{equation}
This scalar value must be averaged over the path, because
the average force is the work done by it during the jump divided by the
length of the path.
By this a frequently used assumption, in which the wind force
is taken as the average of its values at the initial and saddle point
position \cite{Gup82}, can be tested.

\section{Results for FCC metals}
\label{sec:results}

The formalism described above is applied to metals with the close-packed
face-centered cubic (FCC) and hexagonal close-packed (HCP) structures,
the latter being replaced by its equivalent FCC structure.
Wind valences are calculated for diffusing
atoms in aluminum, the noble metals and
the $4d$ transition metals and are compared to experimental data
and results from previously published computational studies.

\subsection{Electromigration in aluminum}
\label{subsec:Al}

\subsubsection{Self-electromigration}
\label{sssec:selfAl}

\begin{figure}[ht]
\epsfig{figure=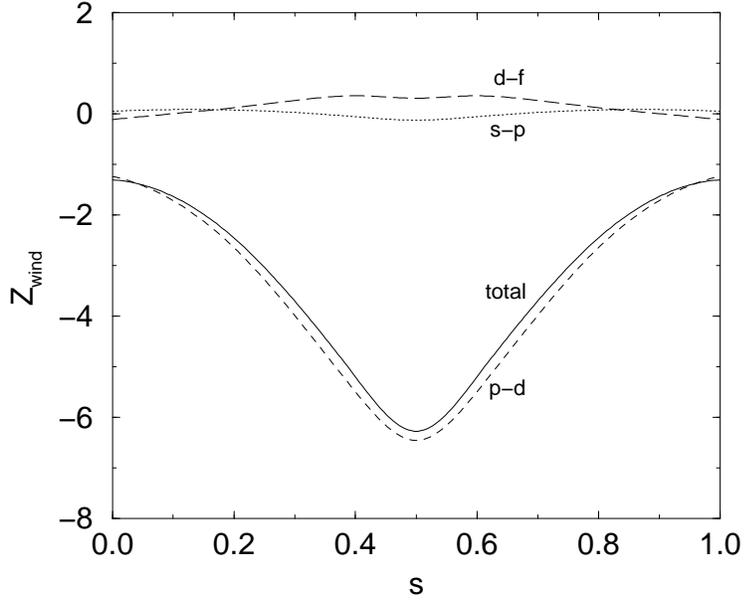,height=9.0cm}
\caption{The variation of the wind valence of a migrating host atom in
aluminum at 800 K.
The $sp$, $pd$ and $df$ contributions are given separately.}
\label{fig:Alself}
\end{figure}

In Fig. \ref{fig:Alself} the solid line shows the variation of the
scalar wind valence of a host atom in aluminum metal
along the migration path.
In addition the three partial wave contributions to $Z_{{\rm wind}}$
are shown,
namely the $sp$, $pd$ and $df$ contributions, depending on the differences
$\eta_{l+1}-\eta_{l}$ as they appear in the expression for the
matrix element
$<\Psi_{k}|-\nabla_{{\bf R}_{p}}v_{p}|\Psi_{k}>$ in Eq.
(\ref{eq:expval4}).
The $pd$ term is dominant which is consistent with
the values of the phase shifts of the moving aluminum atom,
being 0.338, 0.395, 0.051 and 0.002 for $s$, $p$,
$d$ and $f$ respectively at the initial position and
0.370,  0.430, 0.056 and 0.002 at the saddle point.
A sinusoidal behaviour of the wind valence along the path would justify
the averaging procedure, which uses the values in initial and saddle
point position only.
Such a behaviour seems to be obeyed in the figure.
In Table
\ref{table:selfAl} results for the initial and saddle point
position, their average value and the average over the path
are listed.
In the first three columns the three partial contributions are shown.
The fourth contains the sum of the three.
In order to obtain these values the transport relaxation time of
$\tau=69 \ {\rm a.u.}$ has
been calculated using Eq. (\ref{eq:tau}), assuming the resistivity
to equal the phonon part at 800 K, namely $\rho = 8.6 \ \mu \Omega {\rm cm}$
\cite{LanBor82}.
Comparing the third and fourth row one sees that the two point average
$Z_{\rm wind}$ differs 10 \% from the path average value.
In all tables values for the quantity $K$, defined in Eq. (\ref{eq:K}),
and $Z_{{\rm wind}}/\tau$ are given too.
The advantage of specifying $K$ is its weak temperature dependence
(only broadening of the Fermi-Dirac distribution),
while $Z_{\rm wind}$ strongly
depends on temperature through $\rho$.
The variation of $K$ turns out to be less than 1 \% in a
temperature range of 1000 K for aluminum,
so $K$ can be considered constant and is shown in the sixth column of
Table \ref{table:selfAl}.
The theoretically purest quantity to be extracted from our calculation is
$Z_{\rm wind}/\tau$, shown in the last column of this table.
The calculation of this quantity requires no model for the resistivity
whatsoever.
A disadvantage of this quantity is the difficulty in comparing it with
experimental quantities.

Very recently Ernst et al. \cite{ErnFroWev96a} presented experimental
results for bulk electromigration in aluminum.
They found an effective valence varying from -5 at 683 K to -3.3 at
883 K.
When we use the combination of the Eqs. (\ref{eq:Ftot}) and (\ref{eq:K})
\begin{equation}
\label{eq:effZ}
Z^{*}(T)=Z_{\rm direct}+\frac{K}{\rho(T)}
\end{equation}
in order to analyse these measurements, this leads to values of
$K=-45 \ \mu \Omega {\rm cm}$ and
$Z_{\rm direct}=+1.4$.
The value of $K$ is about 1.5 times larger than the calculated value of 
$-29 \ \mu \Omega {\rm cm}$ (see
Table \ref{table:selfAl}), which is acceptable.
It is interesting that the direct valence is very close
to the value of half the chemical
valence as predicted by Bosvieux and Friedel \cite{BosFri62}.

Sorbello \cite{Sor73} has performed model-pseudopotential
calculations for self-electromigration in Al and found
a $K$ value of $-112 \ \mu \Omega {\rm cm}$.
Although a pseudopotential formulation is considered to be suitable
for a nearly free-electron system like aluminum, his value is rather
different
from our calculated value and the experimental value found by
Ernst et al. \cite{ErnFroWev96a}.
On the other hand his value agrees approximately with values given by
Lodding \cite{Lod64},
who refers to measurements by Penney \cite{Pen64}.
With respect to these experiments we point out, that
all available theoretical models for bulk electromigration lead to a
temperature dependence of the effective valence according to
Eq. (\ref{eq:effZ}).
The measurements of Penney cannot be fitted by this formula.

An interesting phenomenon is the influence of impurities on the
electromigration properties of aluminum.
The best-known example is the addition of small amounts of copper,
which reduces electromigration induced damage effectively.
In the hope that other impurities have a similar but stronger
effect on electromigration in aluminum,
experimentalists tried palladium and silicon on an ad-hoc basis,
but without success.
In order to contribute to a microscopic explanation we investigated the
influence of copper, palladium and silicon impurities, that
are located in the impurity cluster
near a migrating aluminum atom \cite{LodDekAth96}.
However, we found \cite{DekLod96} that on average the effect is small,
although the presence of impurity atoms at particular positions can give
rise to a considerable reduction of the wind valence.
The presence of a palladium atom induces a reduction of
about 10 \% when averaged
over all positions neighbouring the migration path (shaded atoms in Fig.
\ref{fig:void}).
The effect of silicon and copper is even smaller.

The presence of impurities changes the number of valence
electrons in the system.
Within a rigid band model this corresponds to a change of the Fermi
energy.
A vacancy or a copper atom lowers the Fermi level, while
silicon raises $\epsilon_{F}$.
The addition of 1 \% of copper reduces the average number of valence
electrons per atom from 3.0 to 2.98.
According to our calculations \cite{DekLod96}
this reduces the wind valence only by a few percent.
The conclusion is, that copper atoms do not reduce the wind valence
of aluminum atoms directly through electronic effects.

Finally we mention the effect of impurities due to their
contribution to the electrical resistivity,
which causes a decrease of the wind valence.
An addition of 1 \% of copper induces a resistivity increase of
about 10 \% and
the wind valence will be reduced by about 10 \%.

We conclude that all impurity effects considered above do not induce a
dramatic change in the value of the wind valence.
Therefore the reduction of the
electromigration induced damage cannot be attributed to them.

\subsubsection{Impurities in aluminum}
\label{sssec:impAl}

\begin{figure}[ht]
\epsfig{figure=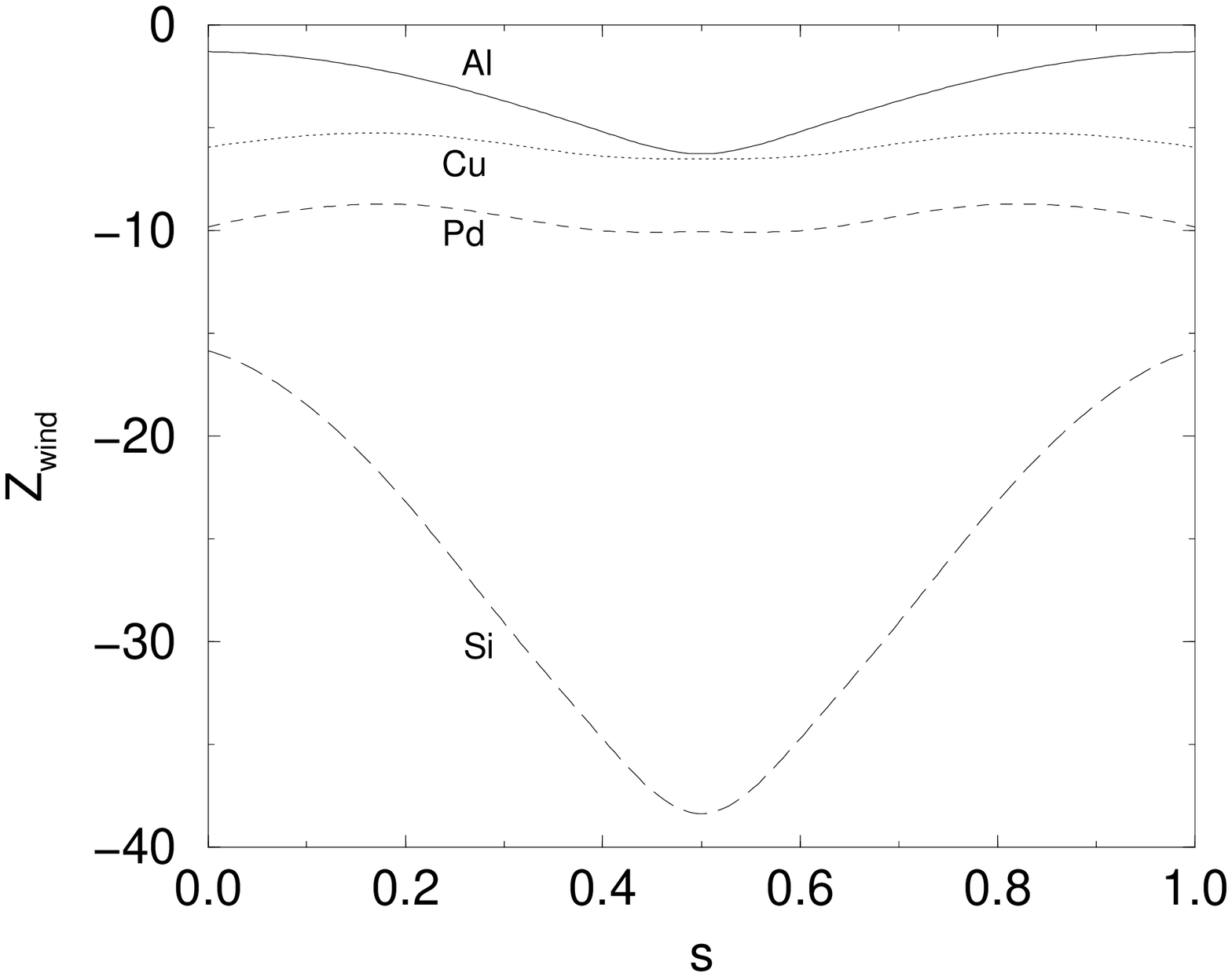,height=9.0cm}
\caption{The position dependent wind valence of 
copper, palladium and silicon impurities in Al.}
\label{fig:impAl}
\end{figure}

Besides effects on the electromigration properties
of host atoms, impurity atoms can electromigrate themselves.
In principle, impurities can migrate in a direction opposite to host
atom transport.
However, the path dependent
wind valences of copper, palladium and silicon, shown in Fig.
\ref{fig:impAl}, all have a negative sign and are larger in
magnitude than $Z_{{\rm wind}}$ of aluminum,
which is shown for comparison.
This rather accelerates the vacancy transport than slowing it down.
The average wind valences of copper, silicon and palladium are
factors of
1.7, 2.8 and 7.8, respectively, larger than $Z_{{\rm wind}}$ of aluminum.
This is in partial contradiction with
measurements, quoted by Ho and Kwok \cite{HoKwo89} in their
review article, giving copper a smaller wind valence than host aluminum.
We note that the calculated ratios are independent of the model for the
calculation of $\tau$, because $\tau$ is a host quantity.
The fact that the wind valences of the impurities have the same sign is
not surprising, because the direction of the force is mainly
determined by the direction of the motion of the charge carriers.
It is worth noting, that the assumption of a sinusoidal variation
of the wind valence along the path does not hold
both for copper and palladium, which can be seen clearly
in Fig. \ref{fig:impAl}.
This makes that in general
the average using only the initial and saddle point position
values is not a good measure for the average over the entire path.

\subsection{Electromigration in the noble metals}
\label{subsec:noble}

In this subsection self-electromigration in the noble metals copper,
silver and gold is studied as well as impurity migration in silver.

\subsubsection{Self-electromigration in copper, silver and gold}
\label{sssec:nobleself}

\begin{figure}[ht]
\epsfig{figure=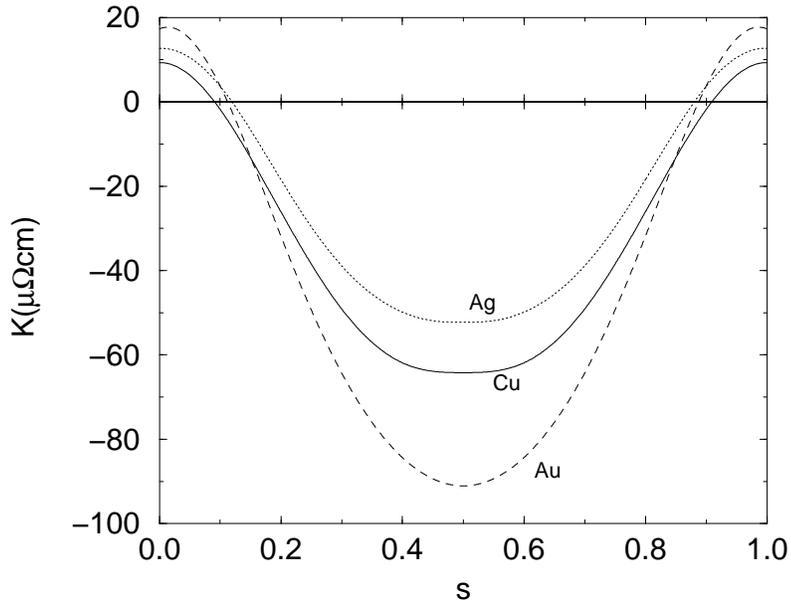,height=9.0cm}
\caption{Variation of $K$ along the path for self-electromigration in
Cu, Ag and Au.}
\label{fig:nobself}
\end{figure}

The variation of $K$ along the path for self-electromigration in
the noble metals copper (solid line), silver (dotted) and gold (dashed)
is shown in Fig.
\ref{fig:nobself}.
A sign change from positive at the initial site to negative further
down the path is observed.
This cannot be understood in terms of electron and hole conduction
and a free-electron-like model is not able to reproduce
such behaviour.

The calculated $K$ values listed in Table \ref{table:wijGupSor}
cannot be compared with experiment, because as far as a
temperature dependence of the effective valence has been
measured \cite{DoaBre70,GilLaz66}, the results
cannot be interpreted in terms of a constant $K$.
But it is interesting to compare them to
$K$ values calculated by Gupta \cite{Gup82} and Sorbello \cite{Sor73}.
Our values for copper and silver compare very well with
the ones of Gupta, while our value for gold is a factor of two smaller.
The values of Sorbello are very different from our results,
which is not surprising in view of the pseudopotential method he used.
However, similar trends are found in our and in Sorbello's calculations:
gold shows the largest value of $K$, while silver shows the smallest.
In the calculation of Gupta the $K$ value of copper is the smallest.

In Table \ref{table:nobself} we give calculated
$Z_{\rm wind}$ and measured $Z^{*}$ values
\cite{DoaBre70,GilLaz66,Sul67,PatHun70} at a given temperature
$T$, while
the last column contains the quantity $Z_{\rm wind}/\tau$.
The calculated wind valences are of the same order of magnitude as the
experimental effective valences,
except for the much larger $Z^{*}$ of Doan and Brebec for silver
\cite{DoaBre70}.
In order to make a detailed comparison between theory and experiment
more experimental information is needed.

\subsubsection{Impurities in silver}
\label{sssec:impAg}

\begin{figure}[ht]
\epsfig{figure=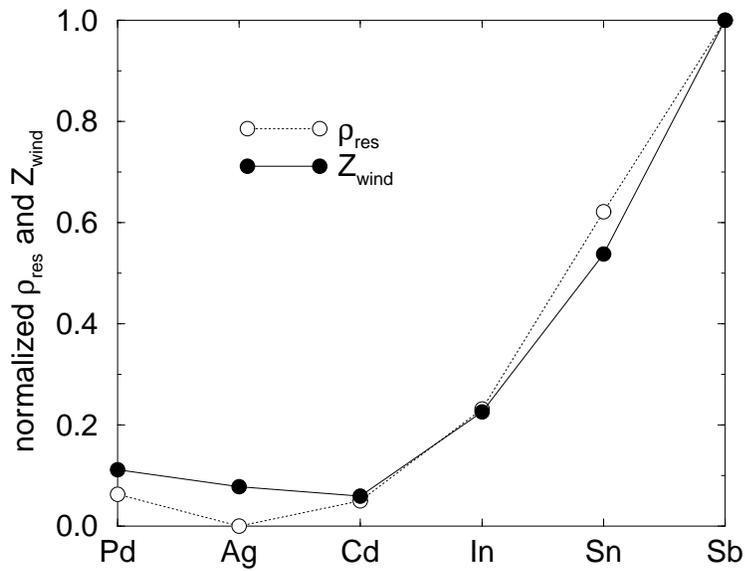,height=9.0cm}
\caption{Measured residual resistivity and calculated wind valence of
Pd, Ag, Cd,
In, Sn and Sb impurities in Ag. Both curves are normalized 
with respect to the Sb data.}
\label{fig:ZresAg}
\end{figure}

This section is devoted to the wind valence of
palladium, silver,
cadmium, indium, tin and antimony impurities in silver.
The series is part of row 5 in the Periodic System.
The calculated wind valence essentially follows the trend of the
experimental residual resistivity \cite{LanBor82} as is shown in
Fig. \ref{fig:ZresAg}, where both quantities are given as a
function of the chemical valence.
Both quantities are normalized with respect to the largest values,
which occur for antimony.
Note that the wind valence of the cadmium impurity is smaller than
the one of the host atom.

\begin{figure}[ht]
\epsfig{figure=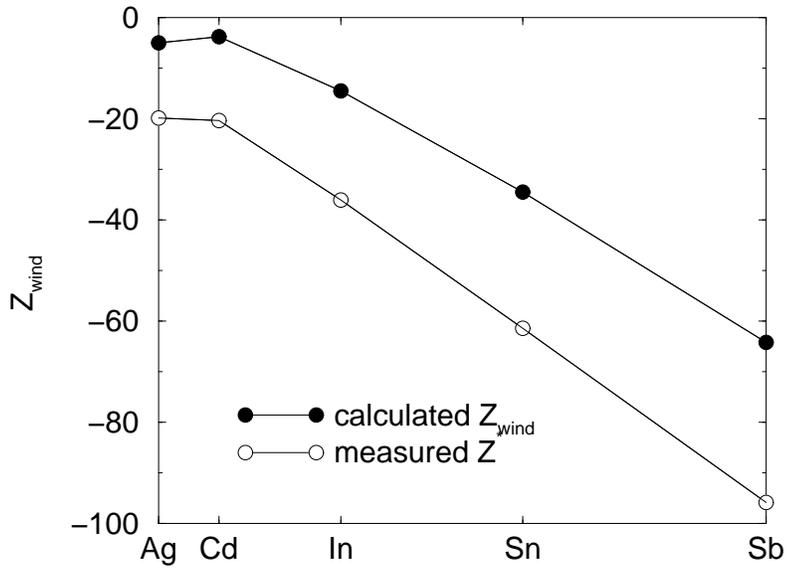,height=9.0cm}
\caption{The solid circles give the calculated wind valence of the
atoms Ag, Cd, In, Sn and Sb in Ag
as a function of $z(z+1)$, where $z=Z_{\rm atom}-Z_{\rm Ag}$.
The open circles are effective valences, measured by Doan
{\protect \cite{Doa70}} at temperatures of
approximately 1150 K.}
\label{fig:zzplus1}
\end{figure}

With the exception of palladium,
Doan \cite{Doa70} investigated this series of
impurities experimentally and found the
effective valence to be linear with $z(z+1)$, where
$z=Z_{\rm imp}-Z_{\rm Ag}$, with $Z_{\rm imp}$ and $Z_{\rm Ag}$ the chemical
valence of the impurity and silver atoms respectively.
This dependence is a result of
Mott's theory for Born scattering, which leads
to a wind valence of an impurity atom
proportional to $z^2$ at its initial position (neglecting backscattering
from the vacancy) and to $Z_{\rm imp}^2=(z+1)^2$ at the saddle point.
The calculated $Z_{{\rm wind}}$ and the measured $Z^{*}$ at about 1150 K
are given as a function of $z(z+1)$ in Fig. \ref{fig:zzplus1}.
Measurements and calculations show a similar trend, but the values do
not agree.
Neither the measured $Z^{*}$ nor the calculated $Z_{{\rm wind}}$
of the silver atom show the $z(z+1)$ dependence.

\begin{figure}[ht]
\epsfig{figure=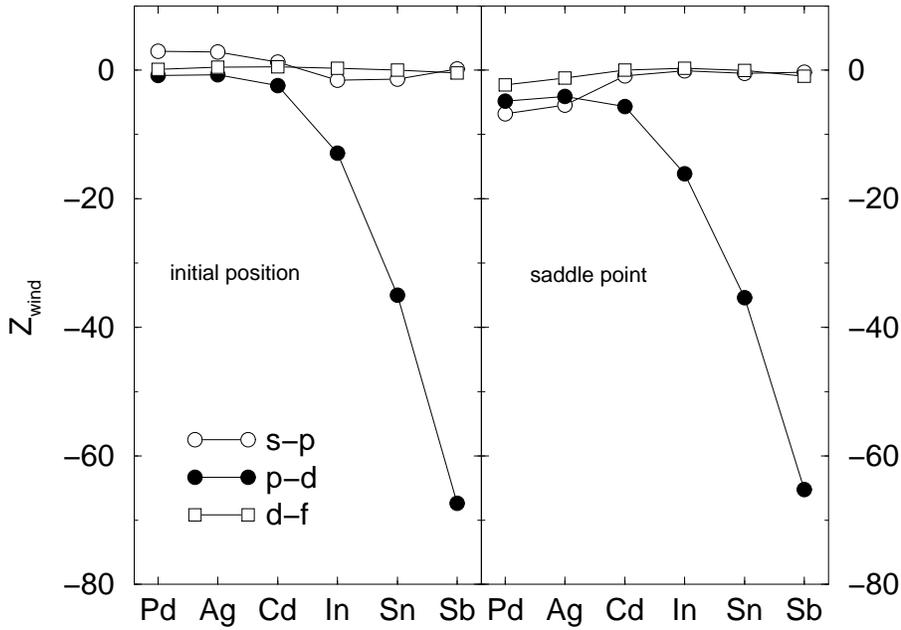,height=9.0cm}
\caption{$sp$, $pd$ and $df$ contributions to the wind valence of
$4d$ and $5sp$ impurities in Ag at $T=800 {\rm K}$.}
\label{fig:impAgspdf}
\end{figure}

More information about the trend in the wind valence can be extracted
by considering the $sp$, $pd$ and $df$ contributions separately.
They are plotted in Fig. \ref{fig:impAgspdf} for the initial and saddle
point position, respectively.
A transport relaxation time of $\tau = 243$ is used,
corresponding to a resistivity of $\rho = 4.9 \ \mu \Omega {\rm cm}$ at
800 K \cite{LanBor82}.
The $pd$ term turns out to determine the overall behaviour.
It develops as soon as the $p$ states of the impurity atom are occupied,
starting with indium.
Note that for the saddle point configuration the $sp$ term vanishes
when the $5s$ orbit is completely filled,
at cadmium, and leads to a relatively low
wind valence of cadmium.
A possible origin of it can be the following.
In the saddle point configuration the impurity replaces two host
atoms.
For an impurity atom with filled $5s$ orbitals, two $5s$ electrons
are present, just as in the configuration with host atoms in initial and
final positions, which is just the host configuration.
Such an impurity at the saddle point resembles an unperturbed host
state as far as the $s$-electrons are concerned.
It may be the case that scattering involving these orbitals is
weaker than when just one $5s$ electron is present.

\subsection{Self-electromigration of the $4d$ transition metals}
\label{subsec:4dtrans}

We also applied the theory to the $4d$
transition metals, mainly as a challenge to experimentalists.
To our knowledge only electrotransport in
zirconium has been measured \cite{HoKwo89},
showing a small positive effective valence of 0.3.
A small positive valence has also been measured for the $5d$
transition metal platinum.
Only metals with a close-packed structure are considered in this series
of calculations.
Metals with the hexagonal close-packed (HCP) structure, namely
yttrium, zirconium,
technetium and ruthenium, are treated as FCC metals.

\begin{figure}[ht]
\epsfig{figure=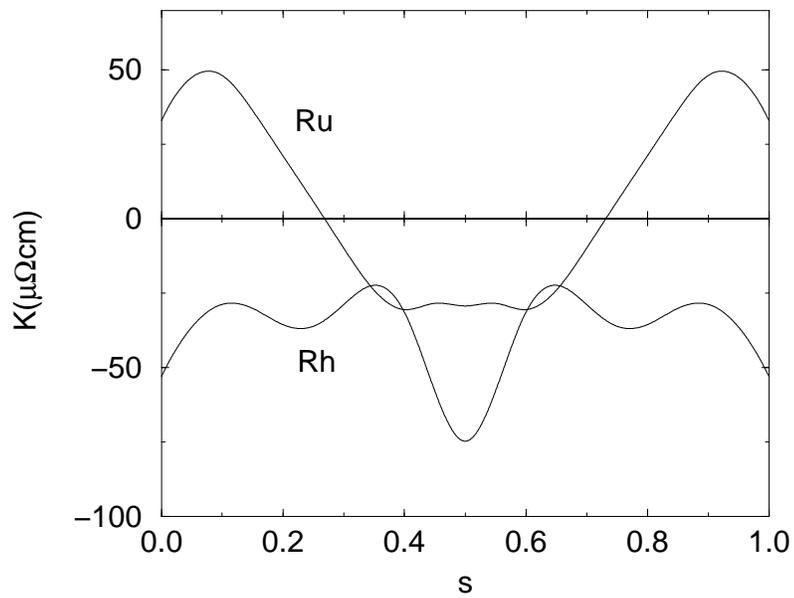,height=9.0cm}
\caption{Variation of $K$ along the migration path for Rh and Ru
self-electromigration.}
\label{fig:RhRu}
\end{figure}

Remarkable variations of $K$ along the path occur, as shown for
rhodium and ruthenium in Fig. \ref{fig:RhRu}.
One observes minima, maxima and sign changes.
Simple models will fail to describe such features.
It should be stressed that the variations occur in spite of the smoothly
varying phase shifts, so they are likely to occur due to multiple scattering
effects.

The $K$ values averaged over the path are given in Table \ref{table:4d}.
In our procedure
for the construction of the potentials the atomic configuration
is important.
The electronic configuration of a free atom sometimes differs from that
of an atom embedded in a crystal.
This is the case for palladium, which has a $4d^{9}5s^{1}$ configuration
in the crystal, while the free atom has a $4d^{10}5s^{0}$ configuration.
The electronic configuration of a host atom in a rhodium crystal
lies somewhere between the
$4d^{8}5s^{1}$ and $4d^{9}5s^{0}$ configurations.
As can be seen from the table the value of $K$ depends only slightly
on the configuration used.
So the wind valence appears to be only weakly dependent on the
precise electronic structure obtained from non-selfconsistent
potentials.
Although it remains interesting to investigate the influence of
self-consistency we do not expect large effects for a metal like
rhodium.
Looking along the series in Table \ref{table:4d} no clear trend
is observed in the $K$ values.

Also shown are the wind valences at a certain (high) temperature
and the transport relaxation time $\tau$ used to calculate it.
The effective valence of zirconium, which was measured to be +0.3,
is not in contradiction with the value of -0.4 for
the wind valence
However, it should be mentioned that the
experiments were done at high temperatures in the $\alpha$-phase
of zirconium.
The wind valence for this phase, having the BCC structure,
will be calculated in the near future.
As can be seen in the table, the
wind valence in the $4d$ transition metals turns out to be small.
On the other hand the chemical valence can be rather large and
therefore the effective valence
is dominated by the direct valence and varies only slightly with
temperature.
Hence, these transition metals are suitable for the experimental
determination of
the direct valence, which has been the subject of a long-lasting
controversy \cite{Lod91clar}.

\section{Conclusions}
\label{sec:conclusions}

We have improved a Green's function method for the
calculation of the electronic structure in dilute alloys.
The formalism has been applied to the calculation
of the wind force in the case of substitutional electromigration.
We have focussed on FCC and HCP metals with
the HCP metals treated in the FCC structure.

The calculated wind valence for self-electromigration in aluminum is
in acceptable
agreement with recent measurements of Ernst et al. \cite{ErnFroWev96a}.
We have also investigated the effect of the presence of
impurities on
the wind valence of aluminum.
Neither the presence of an impurity atom near the jump path nor
the impurity induced increase of the total resistivity of aluminum
induces a dramatic change of the wind valence.
So, changes in the electronic structure due to the presence of
an impurity are not the reason for a reduction of $Z_{\rm wind}$.

The wind valence for self-electromigration in the noble metals
shows a sign change along the path.
Such a behaviour cannot be reproduced by simple models.
Path averaged values show qualitative agreement with experiment.

Calculated wind valences of cadmium, indium, tin and antimony
in silver roughly follow the measured residual resistivity
and are approximately linear with $z(z+1)$, where $z$ is 
the difference in chemical valence between the impurity and the
silver.
Such a trend was predicted
within Mott's theory for Born scattering in metals
and has been observed experimentally, although
the precise values of the wind valence do not agree very well.

Finally the wind valences for self-electromigration in the $4d$
transition metals show a large variation in size and sign.
The average values often are small due to cancellation of the wind valence
along the migration path.
The effective valence therefore is dominated by the
direct valence and will depend on temperature only weakly.
Hence, measurements on these metals are suitable to decide which
model for the direct force is the correct one.

\section*{acknowledgement}

This work was sponsored by the Stichting Nationale Computerfaciliteiten
(National Computing Facilities Foundation, NCF) for the use of
supercomputer facilities, with financial support from the Nederlandse
Organisatie voor Wetenschappelijk Onderzoek (Netherlands Organization
for Scientific Research, NWO).

\appendix
\section{}
\label{app:B}

First Eq. (\ref{eq:coefp}) will be derived from Eq. (\ref{eq:GvE}).
In Eq. (\ref{eq:coefp}), written in the position representation,
one substitutes the local Schr\"odinger equation
\begin{equation}
v^{p}({\bf x})\Psi_{k}({\bf x}+{\bf R}_{p})=
(\nabla^2 +E)\Psi_{k}({\bf x}+{\bf R}_{p}).
\label{eq:SEPsi}
\end{equation}
Then one applies Green's theorem and uses the equation for the void
Green's function
\begin{equation}
\label{eq:Gvdelta}
(\nabla^2 +E)G^{{\rm void}}({\bf r},{\bf r}')=\delta({\bf r}-{\bf r}'),
\end{equation}
as it holds inside the void region.
By that the void wave
function can be written as a sum of spherical surface integrals
\begin{eqnarray}
\Psi^{\rm void}_{k}({\bf x}+{\bf R}_{p})=
&&\sum_{p'} \int_{S_{p'}} d{\bf S}_{p'}\cdot 
(\nabla' G^{{\rm void}}({\bf x}+{\bf R}_{p},{\bf x}'+{\bf R}_{p'}))
\Psi_{k}({\bf x}'+{\bf R}_{p'})-\nonumber \\
&&G^{{\rm void}}({\bf x}+{\bf R}_{p},{\bf x}'+{\bf R}_{p'})
\nabla' \Psi_{k}({\bf x}'+{\bf R}_{p'}).
\label{eq:Surfint}
\end{eqnarray}
Using Eq. (\ref{eq:Gvoid}) for the void Green's function and the form of the
local basis functions at the boundaries of the cell in Eq. (\ref{eq:RLp})
Eq. (\ref{eq:Surfint})
can be elaborated for small $x$. This leads to the expression
(\ref{eq:coefp}) for the void wave function coefficients of Eq.
(\ref{eq:Psivoid}) in terms
of the alloy wave function coefficients of Eq. (\ref{eq:localPsi}).

Now we want to illustrate the derivation of Eq. (\ref{eq:cvjcvp}), the
relation between $c^{{\rm void}}_{kpL}$ and $c^{{\rm void}}_{kjL}$.
This can be found by working out the void wave function for a position
${\bf r}={\bf x}+{\bf R}_{j}={\bf x}_{p}+{\bf R}_{p}$, which is a
position in the alloy cell labeled by $p$ and in the host cell
labeled by $j$,
\begin{equation}
\label{eq:Psivoid1}
\Psi^{\rm void}_{k}({\bf r})=\sum_{L}c^{\rm void}_{kpL}j_{L}({\bf x}_{p})=
\sum_{L}c^{\rm void}_{kjL}j_{L}({\bf x}).
\end{equation}
The two Bessel functions centered at
different positions are related by\begin{equation}
j_{L}({\bf x})=j_{L}({\bf x}_{p}+{\bf R}_{pj})
=\sum_{L'}j_{L'}({\bf x}_{p})J^{pj}_{L'L},
\label{eq:Jpj}
\end{equation}
in which the matrix $J^{pj}$ is given by
\begin{equation}
\label{eq:Jnn'}
J^{pj}_{LL'}=4 \pi i^{l-l'}\sum_{L''}i^{l''}C_{LL'L''}j_{L''}({\bf R}_{pj}).
\end{equation}
Substitution of this equation in Eq. (\ref{eq:Psivoid1}) automatically leads to
Eq. (\ref{eq:cvjcvp}).

\section{}
\label{app:C}

It will be shown how the infinite angular momentum summation in Eq.
(\ref{eq:coef}) can be carried out analytically, leading to
Eq. (\ref{eq:coef1}).
In addition 
Eq. (\ref{eq:GVisGminGAG2}) is generalized to Eq. (\ref{eq:GV}), valid for
arbitrary position labels.

First a relation between ${\cal G}^{{\rm void},pp'}$, present in
the left hand side of Eq. (\ref{eq:coef}), and
${\cal G}^{{\rm void},jj'}$ will be derived.
To that end the void Green's function $G^{{\rm void}}({\bf r},{\bf r}')$
for the positions
${\bf r}={\bf x}+{\bf R}_{j}={\bf x}_{p}+{\bf R}_{p}$ and
${\bf r}'={\bf x}'+{\bf R}_{j'}={\bf x}'_{p'}+{\bf R}_{p'}$,
see Eq. (\ref{eq:Gvoid}),
is rewritten using the free-space Green's function (\ref{eq:G0}).
It follows that
\begin{eqnarray}
\label{eq:Gvoid1}
G^{{\rm void}}({\bf r},{\bf r}')&&=\nonumber \\
G^{0}({\bf r},{\bf r}')
&&+\sum_{LL'}j_{L}({\bf x}_{p})
({\cal G}^{{\rm void},pp'}_{LL'}-B^{pp'}_{LL'})j_{L}({\bf x}'_{p'})=\nonumber \\
G^{0}({\bf r},{\bf r}')
&&+\sum_{LL'}j_{L}({\bf x})
({\cal G}^{{\rm void},jj'}_{LL'}-B^{jj'}_{LL'})j_{L}({\bf x}').
\end{eqnarray}
Applying Eq. (\ref{eq:Gvoid1}) and using Eq. (\ref{eq:Jpj}) straightforwardly
leads to
\begin{equation}
\label{eq:secTermGv}
{\cal G}^{{\rm void},pp'}=B^{pp'}+
J^{pj}({\cal G}^{{\rm void},jj'}-B^{jj'})J^{j'p'}.
\end{equation}
Now the right hand side of Eq. (\ref{eq:coef}) is rewritten using
the KKR equation (\ref{eq:KKRc}) and Eq. (\ref{eq:KKRmat})
\begin{equation}
\label{eq:RHScoef}
\sum_{j'}J^{pj}
(1-{\cal G}^{{\rm void}}t^{{\rm h}})^{jj'}
c^{{\rm host}}_{kj'}=
J^{pj}b({\bf k})t^{{\rm h}}c^{{\rm host}}_{kj}
-\sum_{j'}J^{pj}{\cal G}^{{\rm void},jj'}t^{{\rm h}}c^{{\rm host}}_{kj'}.
\end{equation}
The first term on the right hand side of this equation can be evaluated
using the expansion
\begin{equation}
\label{eq:JB}
B^{pj'}=J^{pj}B^{jj'},\hspace{2cm}(R_{pj}<R_{jj'})
\end{equation}
which can be derived from an expansion for the
Hankel function similar to Eq. (\ref{eq:Jpj}) for the Bessel function.
From the definition of $b({\bf k})\equiv b({\bf k},0)$,
see Eq. (\ref{eq:bkp}), it follows that
\begin{equation}
J^{pj}b({\bf k})=b({\bf k},{\bf R}_{p})e^{i{\bf k}\cdot{\bf R}_{pj}}-
B^{pj}.
\label{eq:Jbk}
\end{equation}
Substituting this equation in Eq. (\ref{eq:RHScoef}) the first term on the
right hand side of Eq. (\ref{eq:coef1}) appears, and one has to deal
further with the terms
\begin{equation}
-\sum_{j'}(B^{pj'}\delta_{jj'}+
J^{pj}{\cal G}^{{\rm void},jj'})t^{{\rm h}}c^{{\rm host}}_{kj'}.
\label{eq:1TRHS}
\end{equation}
The sum of matrices
$B^{pj'}\delta_{jj'}+J^{pj}{\cal G}^{{\rm void},jj'}$ can be written
as $B^{pj'}+J^{pj}({\cal G}^{{\rm void},jj'}-B^{jj'})$.
The latter sum is equal to ${\cal G}^{{\rm void},pj'}$, given by
Eq. (\ref{eq:secTermGv}) with $p'$ replaced by a host label $j'$.
By this Eq. (\ref{eq:RHScoef}) obtains the form of Eq. (\ref{eq:coef1}).

Now we turn to the derivation of Eq. (\ref{eq:GV}).
Substitution of Eq. (\ref{eq:GVisGminGAG2}) in Eq. (\ref{eq:secTermGv}) yields
\begin{equation}
\label{eq:secTermGv2}
{\cal G}^{{\rm void},pp'}=B^{pp'}+
J^{pj}({\cal G}^{jj'}-B^{jj'})J^{j'p'}
-\sum_{j_{1}j_{2}}J^{pj}{\cal G}^{jj_{1}}
{({t^{{\rm h}}}^{-1}+{\cal G})}^{-1}_{j_{1}j_{2}}
{\cal G}^{j_{2}j'}J^{j'p'}.
\end{equation}
Regarding Eq. (\ref{eq:calG}) for ${\cal G}^{jj'}$ one has to apply 
Eq. (\ref{eq:Jbk}) for a further reduction.
Using the straightforward generalization of the host Green's function
matrix (\ref{eq:calG}), defined by
\begin{equation}
\label{eq:calGnnapp}
{\cal G}^{pp'}\equiv \frac{1}{\Omega_{BZ}}\int_{BZ}d^{3}k
[b({\bf k},{\bf R}_{pp'})+b({\bf k},{\bf R}_{p})M^{-1}({\bf k})
b^{{\rm T}}(-{\bf k},{\bf R}_{p'})]e^{i{\bf k}\cdot{\bf R}_{pp'}},
\end{equation}
the following relations can be derived
\begin{mathletters}
\label{eq:JcalGJ}
\begin{eqnarray}
\label{eq:JcalGBJ}
J^{pj}({\cal G}^{jj'}-B^{jj'})J^{j'p'}=&&{\cal G}^{pp'}-B^{pp'}
-{\cal G}^{pj'}t^{{\rm h}}B^{j'p'}-B^{pj}t^{{\rm h}}{\cal G}^{jp'}\nonumber \\
&&+B^{pj}t^{{\rm h}}({t^{{\rm h}}}^{-1}+{\cal G})^{jj'}t^{{\rm h}}B^{j'p'}
\end{eqnarray}
\begin{equation}
\label{eq:JcalG}
J^{pj}{\cal G}^{jj_{1}}={\cal G}^{pj_{1}}
-B^{pj}t^{{\rm h}}({t^{{\rm h}}}^{-1}+{\cal G})^{jj_{1}}
\end{equation}
\begin{equation}
\label{eq:calGJ}
{\cal G}^{j_{2}j'}J^{j'p'}={\cal G}^{j_{2}p'}
-({t^{{\rm h}}}^{-1}+{\cal G})^{j_{2}j'}t^{{\rm h}}B^{j'p'}.
\end{equation}
\end{mathletters}
Substituting these equations in Eq. (\ref{eq:secTermGv2})
gives the desired form (\ref{eq:GV})
\begin{equation}
\label{eq:GVapp}
{\cal G}^{{\rm void},pp'}={\cal G}^{pp'}-
\sum_{j_{1}j_{2}}{\cal G}^{pj_{1}}
{({t^{{\rm h}}}^{-1}+{\cal G})}^{-1}_{j_{1}j_{2}}
{\cal G}^{j_{2}p'}.
\end{equation}


\newpage

\begin{thebibliography}{10}

\bibitem{CoeSpe30}
A. Coehn and W. Specht, Z. Physik {\bf 62},  1  (1930).

\bibitem{SorLodHov82}
R.~S. Sorbello, A. Lodder, and S.~J. Hoving, Phys. Rev. B {\bf 25},  6178
  (1982).

\bibitem{HoKwo89}
P.~S. Ho and T. Kwok, Rep. Progr. Phys. {\bf 52},  301  (1989).

\bibitem{Lod91clar}
A. Lodder, Solid State Comm. {\bf 79},  143  (1991).

\bibitem{k}
J. Korringa, Physica {\bf 13},  392  (1947).

\bibitem{kr}
W. Kohn and N. Rostoker, Phys. Rev. {\bf 94},  1111  (1954).

\bibitem{Lod76}
A. Lodder, J. Phys. F: Met. Phys. {\bf 6},  1885  (1976).

\bibitem{1984}
A. Lodder, J. Phys. F: Met. Phys. {\bf 14},  2943  (1984).

\bibitem{VEkDekLod95}
J. van Ek, J.~P. Dekker, and A. Lodder, Phys. Rev. B {\bf 52},  8794  (1995).

\bibitem{OppLod87I}
P.~M. Oppeneer and A. Lodder, J. Phys. F: Met. Phys. {\bf 17},  1885  (1987).

\bibitem{EkLod91I}
J. van Ek and A. Lodder, J. Phys. Cond. Mat. {\bf 3},  7307  (1991).

\bibitem{BosFri62}
C. Bosvieux and J. Friedel, J. Phys. Chem. Solids {\bf 23},  123  (1962).

\bibitem{Nes90}
R.~K. Nesbet, Phys. Rev. B {\bf 41},  4948  (1990).

\bibitem{Zim72}
J.~M. Ziman,  in {\em Principles of the theory of solids}, edited by J.~M.
  Ziman (Cambridge University Press, Cambridge, 1972).

\bibitem{Fik59}
V.~B. Fiks, Sov. Phys.-Solid State {\bf 1},  14  (1959).

\bibitem{HunGro61}
H.~B. Huntington and A.~R. Grone, J. Phys. Chem. Solids {\bf 20},  76  (1961).

\bibitem{KumSor75}
P. Kumar and R.~S. Sorbello, Thin Solid Films {\bf 25},  25  (1975).

\bibitem{EkLod91II}
J. van Ek and A. Lodder, J. Phys. Cond. Mat. {\bf 3},  7331  (1991).

\bibitem{MolLodCol83}
J. Molenaar, A. Lodder, and P.~T. Coleridge, J. Phys. F: Met. Phys. {\bf 13},
  839  (1983).

\bibitem{OppLod87II}
P.~M. Oppeneer and A. Lodder, J. Phys. F: Met. Phys. {\bf 17},  1901  (1987).

\bibitem{LodBra94}
A. Lodder and P.~J. Braspenning, Phys. Rev. B {\bf 49},  10215  (1994).

\bibitem{LasSov73}
R.~H. Lasseter and P. Soven, Phys. Rev. B {\bf 8},  2476  (1973).

\bibitem{Gup82}
R.~P. Gupta, Phys. Rev. B {\bf 25},  5188  (1982).

\bibitem{LanBor82}
J. Bass,  in {\em Landolt-B\"ornstein, Numerical Data and Functional
  Relationships in science and technology, New Series}, edited by K.-H.
  Hellwege and J.~L. Olsen (Springer-Verlag, Berlin, 1982), Vol.~15a.

\bibitem{ErnFroWev96a}
B. Ernst, G. Frohberg, and H. Wever, Def. Diff. Forum {\bf 143-147},  1649
  (1997), proc. Intl. Conf. 'DIMAT-96' (Nordkirchen, Germany, August 1996).

\bibitem{Sor73}
R.~S. Sorbello, J. Phys. Chem. Solids {\bf 34},  937  (1973).

\bibitem{Lod64}
A. Lodding, J. Phys. Chem. Solids {\bf 26},  143  (1964).

\bibitem{Pen64}
R.~V. Penney, J. Phys. Chem. Solids {\bf 25},  335  (1964).

\bibitem{LodDekAth96}
A. Lodder and J.~P. Dekker,  in {\em Proceedings of the First International
  Alloy Conference(Athens, 1996)}, edited by A. Gonis, A. Meike, and P.~E.~A.
  Turchi (Plenum, New York, 1997).

\bibitem{DekLod96}
J.~P. Dekker and A. Lodder, Def. Diff. Forum {\bf 143-147},  1645  (1997),
  proc. Intl. Conf. 'DIMAT-96' (Nordkirchen, Germany, August 1996).

\bibitem{DoaBre70}
N.~V. Doan and G. Brebec, J. Phys. Chem. Solids {\bf 31},  475  (1970).

\bibitem{GilLaz66}
H.~M. Gilder and D. Lazarus, Phys. Rev. {\bf 145},  507  (1966).

\bibitem{Sul67}
G.~A. Sullivan, J. Phys. Chem. Solids {\bf 28},  347  (1967).

\bibitem{PatHun70}
H.~R. Patil and H.~B. Huntington, J. Phys. Chem. Solids {\bf 31},  463  (1970).

\bibitem{Doa70}
N.~V. Doan, J. Phys. Chem. Solids {\bf 31},  2079  (1970).

\end{thebibliography}

\newpage

\begin{table}[hb]
\caption{Results for self-electromigration in Al.
A $\tau$ value of 69 a.u., based on a resistivity
$\rho = 8.6 \ \mu \Omega {\rm cm}$ at 800 K has been used.}
\label{table:selfAl}
\begin{tabular}{ccccccc}
                  & 
$Z_{\rm wind}(sp)$ &$Z_{\rm wind}(pd)$ &$Z_{\rm wind}(df)$ &
$Z_{\rm wind}$ & $K(\ \mu \Omega {\rm cm})$&
 $Z_{\rm wind}/\tau$ \\
\hline
initial position  &
    0.05            &      -1.24         &     -0.12          &
    -1.31           &     -11.3               &    -0.019              \\
saddle point      &
    -0.13           &     -6.45         &     0.31           &
    -6.27           &    -54.0                &    -0.091              \\
two point average      &
    -0.04           &     -2.61         &     0.13           &
    -3.79           &    -32.7                &    -0.055              \\
path average           &
 0.02               &   -3.54            &    0.17            &
        -3.36       &    -28.9                &    -0.051
\end{tabular}
\end{table}

\begin{table}[hb]
\caption{Calculated $K$ values for the noble metals
compared with results by Gupta {\protect \cite{Gup82}}
and Sorbello {\protect \cite{Sor73}}.}
\label{table:wijGupSor}
\begin{tabular}{cccc}
    &present work   & Gupta \cite{Gup82}  &  Sorbello \cite{Sor73}  \\
\hline
 Cu &-33.3       &  -31.5              &     -196                \\
 Ag &-24.7       &  -33.1              &     -171                \\
 Au &-42.6       &  -83.2              &     -229
\end{tabular}
\end{table}

\begin{table}[hb]
\caption{Wind and effective valences for the noble metals.
The $2^{\rm nd}$ and $3^{\rm rd}$ columns contain the calculated wind
valence and the measured effective valence, respectively, at the temperature
shown in the $4^{\rm th}$ column.
The last column contains the quantity
$Z_{\rm wind}({\rm total})/\tau$ in atomic units.}
\label{table:nobself}
\begin{tabular}{ccccc}
   & $Z_{\rm wind}$ & $Z^{*}$               & $T$(K) & $Z_{\rm wind}/\tau$ \\
\hline
Cu &     -3.5       &  -4.3 \cite{Sul67}    &  1300  & -0.028              \\
Ag &     -3.3       & -19.9 \cite{DoaBre70} &  1150  & -0.021              \\
   &                &  -5.1 \cite{PatHun70} &        &                     \\
Au &     -3.4       &  -6.6 \cite{GilLaz66} &  1289  & -0.037
\end{tabular}
\end{table}

\begin{table}[hb]
\caption{The calculated $K$
and $Z_{\rm wind}$ at a temperature $T$ for the $4d$ transition metals.
The transport relaxation
time $\tau$ at that $T$ is given in atomic units.}
\label{table:4d}
\begin{tabular}{ccccc}
      & $K(\mu \Omega {\rm cm})$& $T$(K) & $Z_{\rm wind}$ & $\tau$(a.u.) \\
\hline
Y                  &    93.2    &  1700  &      0.4       &  25.4        \\
Zr                 &   -52.1    &  1700  &     -0.4       &  13.7        \\
Tc                 &    34.8    &  1700  &      0.5       &  12.4        \\
Ru                 &     7.0    &  1500  &      0.2       &  10.6        \\
Rh($4d^{8}5s^{1}$) &   -36.9    &  1700  &     -1.0       &  13.0        \\
Rh($4d^{9}5s^{0}$) &   -38.3    &  1700  &     -1.0       &  12.7        \\
Pd                 &   -58.6    &  1700  &     -1.3       &  50.4        \\
\end{tabular}
\end{table}

\end{document}